**Artificial Intelligence and the Dual Paradoxes: Examining the Interplay of Efficiency, Resource Consumption, and Labor Dynamics**

**Authors:**

Dr. Mfon Akpan

Assistant Professor of Accounting

Methodist University

Email: makpan@methodist.edu

Dr. Adeyemi Adebayo

Senior Lecturer

University of South Africa (UNISA)

Email: deyemidebayo@zoho.com

# Artificial Intelligence and the Dual Paradoxes: Examining the Interplay of Efficiency, Resource Consumption, and Labor Dynamics

**Abstract**

Artificial Intelligence's (AI) rapid development and growth not only transformed industries but also fired up important debates about its impacts on employment, resource allocation, and the ethics involved in decision-making. It serves to understand how changes within an industry will be able to influence society with that change. Advancing AI technologies will create a dual paradox of efficiency, greater resource consumption, and displacement of traditional labor. In this context, we explore the impact of AI on energy consumption, human labor roles, and hybrid roles widespread human labor replacement. We used mixed methods involving qualitative and quantitative analyses of data identified from various sources. Findings suggest that AI increases energy consumption and has impacted human labor roles to a minimal extent, considering that its applicability is limited to some tasks that require human judgment. In this context, the findings suggest that only 10% of labor roles across major sectors have been impacted by AI.

# 1. Introduction

Artificial Intelligence's (AI) rapid development and growth not only transformed industries but also fired up important debates about its impacts on employment, resource allocation, and the ethics involved in decision-making. It serves to understand how changes within an industry will be able to influence society with that change. Advancing AI technologies will create a dual paradox of efficiency, greater resource consumption, and displacement of traditional labor (Nzubechukwu, et al., 2023). Two paradoxes have been sounded in making sense of artificial intelligence (Jevons Paradox and Moravec's Paradox).

Jevons Paradox, as described by the British economist WS Jevons in the 19th century, means that advances in technology aiming to reduce the use of natural resources can lead to an increased demand for those resources and, hence, negative environmental impacts (Li et al., 2023). These include claiming that if Jevon's paradox is correct, sustainability will be reduced, as it would be impossible to practically achieve reductions to environmental impacts such as carbon emissions (Sedlak, 2023). Such knowledge is helpful for policymakers and companies when considering the Jevon Paradox and the challenge of sustainable development with the potential of artificial intelligence to increase the efficiency of the systems (Gao & Feng, 2023). This paradox targets the ongoing need for a strategy to enhance organizational efficiency and for planning measures concerning organizational sustainability and the markets for human capital.

Challenges in automating sensorimotor tasks appear to have led to the emergence of Moravec's Paradox. Cognitive skills are easy to perform, in contrast to low-level motion control skills, as discovered in Moravec's paradox (Arora, 2023; Rotenberg, 2013). This has raised issues about designing artificial intelligence that imitates the physical actions of humans, hence calling for more innovation on the part of integrated cognitive and motor skills systems (Rotenberg, 2013). As researchers go further in the study of these hybrid systems, incorporating conventional robots with artificial intelligence may open new horizons to foster innovations of automating and symbiosis of humans and robots in different areas (Arora, 2023).

This issue raises the question of the scale of the ecological cost of AI development and implementation and causes a reconsideration of existing practices focusing on technological progress and environmental responsibility (Hao & Demir, 2024). This dual focus encourages responsible innovation and drives the exploration of alternative methodologies that minimize ecological impact while maximizing AI's potential benefits.

There are gaps in automation because of the human sensory-motor skill. These gaps call for the understanding of human capital in relation to integrated self-organization systems and how innovative human abilities that are needed in various applications can be supported, not replaced, by technology. Interfacing between the two is constructive in a manner that opens up the chances of subjecting the work to be done between man and the machine through an optimization of the efforts with a view of overcoming the disadvantages of mechanizing some work totally. Over time, adopting this shared vision will prove imperative for creating application systems that advance efficiency and protect employees' and ecosystems' welfare (Scripter, 2024).

Thus, the significance of learning these paradoxes in AI adoption cannot be underestimated (Autor, 2015; Scott et al., 2024). They can be used to establish the drawbacks and possible future problems in current technology that inform the creation of more efficient means of achieving a relationship between human creativity and mechanical preciseness (Autor, 2015; Scott et al., 2024). Debris, from this knowledge, shapes the ethical considerations of using AI in certain spheres and promotes development that can pave the way to easier-to-use technologies. In this way, addressing these paradoxes creates a vision that researchers and developers incorporate into their work to ensure that the frameworks being established for AI systems are being developed with the right human values (Autor, 2015; Scott et al., 2024).

Therefore, for the reasons highlighted, we will be developing an environment whereby technology that is being incorporated into our day-to-day lives will be responsible. Specifically in this study, we will explore how the Jevon Paradox manifests in AI energy consumption. This is important considering the consequences of increased efficiency in AI technologies may sometimes turn out to be quite surprising, including increased energy use that offsets the gain. The Paradox shows technological development-resource use interdependence to highlight the necessity for sustainable practices in the interest of balancing innovation with environmental stewardship (Shumskaia, 2022). In addition, we will be investigating the persistence of human labor roles due to Moravec's Paradox. Realizing this balance, the imperative of having mechanisms that can limit energy use to a minimum while maximizing AI's potential becomes important. This will mean a multi-dimensional approach to investing in renewable energy sources, optimization of algorithms for efficiency, and embedding a culture of sustainability within the development of tech ((Mohan et al., 2024). We will also analyse the rise of hybrid human-AI roles. These roles increase productivity and tap human intuition and creativity together with machine efficiency, hence creating a synergy that could drive innovation in many industries. According to Bouschery et al. (2023), integrating human-AI collaboration into job landscapes reshapes the way skills and training for the workforce will be conceptualized in light of this new paradigm.

We explore the impact of AI on energy consumption, human labor roles, and hybrid roles widespread human labor replacement. We used mixed methods involving qualitative and quantitative analyses of data identified from various sources. Findings appear to suggest that AI increases energy consumption and has impacted human labor roles to a minimal extent, considering that its applicability is limited to some tasks that require human judgment. In this context, the findings suggest that only 10% of labor roles across major sectors have been impacted by AI.

The rest of the paper is organized thus: Section 2 presents the literature and proposition development. In Section 3, we discuss the methodology before presenting the results in Section 4. Section 5 discusses the findings and the implications of the study for policy and practice. We conclude the study in Section 6.

## 2. Literature review and position development

### 2.1. Artificial intelligence and the dual paradoxes

By fostering collaboration across disciplines, organizations can harness AI's full potential, driving efficiency and innovation that aligns with their strategic goals. It enables two entities to find balance in their implementations of the technology by combining their efforts to formulate effective methods of operation that reflect the code of ethics. Taken together, this comprehensive approach will one-day help organizations to understand and deal with the challenges that arise during AI implementation to change the culture for the continual improvement and learning needed to thrive within an increasingly competitive and fast-paced environment in the future (Weber et al., 2023).

As industries evolve, it is clear that extensive partnerships between educational establishments and employers are going to remain the key to the continuing appropriateness and effectiveness of training programs. By maintaining a linkage system that prioritizes skill enhancement, the two parties can easily solve the shortage of professional abilities and encourage lifelong learning, hence enabling a more robust economy in the next shocks (Yu et al., 2023). Such a dynamic relationship will also promote the sharing of resources and knowledge protection, ensuring that educators and other industry specialists are informed of new trends and technologies in the market.

While AI increases efficiency and, according to the Jevon Paradox, leads to increased consumption of resources and challenging sustainability goals, limitations described by Moravec's Paradox do not allow the complete substitution of human labor by AI on low-level sensorimotor tasks. While this will continue to evolve and include organizations embracing such hybrid roles, the ethical considerations and potential biases of AI systems themselves need consideration so that human judgment remains crucial to the process (Johnson, 2024). It is, therefore, a continuous conversation by stakeholders concerning equity, privacy, and work in an automated world with regard to dynamic interactions between technologies and human work.

**2.1.1. Jevons Paradox:**

Through old and new media, the proposed approach aims at cultivating a more comprehensive appreciation of these dynamics on the part of policymakers and relevant stakeholders so that the promise of technology may empower continued innovations that augment sustainable resource management. This needs to happen in conjunction with multisectoral interfaces that amalgamate knowledge from the environmental sciences, economics, and social justice to design organizational frameworks that prioritize sustainable and enduring gains over short-term and ephemeral benefits (Agbanyo et al., 2023). Another benefit of an interconnected approach is the ability to create a strategy for current and future issues, thus promoting the creation of sustainable communities as well as ecosystems.

Finally, this vision purports to know and embrace change to make societies grow in synergy with their environment and use technology for the enhancement of these societies. Similarity in the modern approach to AI is computational efficiency and scalability issues. Given the ongoing use of AI in organizations to automate jobs, training and development issues arise, especially since AI technologies present Moravec's paradox. Building hybrid jobs that include opportunities for human discretion united with machine productivity requires the development of new types of education that can effectively prepare workers for technical and analytical tasks. For instance, integrating AI literacy into existing curricula will help bring a generation of professionals skilled in navigating complex interactions between humans and machines, ensuring they are not merely passive users but active contributors to technological advancement (Cetindamar et al., 2024).

Furthermore, given the emerging shift towards stronger application of the automated system, it is appropriate to question how individuals with poor access to these resources can be effectively prepared to improve their position in the labor market competition. This focus on inclusive education will be paramount to creating a future where technology enhances the existing human capital and, thus, fosters sustainable growth in view of the sustained changes in economic structures (Akpan, 2024). Following the above discussion, we propose that:

P1: Improvements in AI efficiency (e.g., better algorithms or hardware) result in disproportionate increases in computational energy consumption.

### 2.1.2. Moravec's Paradox:

These issues are gradually being solved with the help of robotics and ML development, and researchers are currently working on implementing sensory feedback and adaptive learning methods to make the job done by AI more natural. With these technological advancements, AI applicability in other areas, including healthcare, manufacturing, and general interaction, enables higher success in delivering its intended functions (Kim et al., 2023). The implementation of AI in these fields becomes a way of maximizing productivity. It introduces new possibilities for further development of those fields, changing the challenges' problem-solving approach and improving the quality of life. Following this discussion, it makes sense to propose that

P2: Sensorimotor tasks requiring nuanced human judgment remain challenging for AI to automate, fully preserving specific human labor roles.

**2.2. Labor Market Dynamics:**

Maximizing the combination of creativity and efficiency will dictate the pace of development and open up opportunities that have not been available before in the field of business (Pavashe et al., 2023). As businesses move to build this type of symbiosis, they need to nurture a culture of learning and growth that allows people to see change as normal and technology as an opportunity. It will be crucial for this culture to be embraced in the subsequent years for organizations to build an adaptable human capital to meet new market conditions and technological discoveries that will help organizations gain sustainable competitive advantage in environments that are characterized by high volatility (Abdulla Essa Al Hebsi & Ali Al-Shami, 2022). These measures will increase satisfaction with individual jobs and promote creativity because they will provide a way of framing problems in more inspiring ways and because people with heterogeneous skills and expertise will be more likely to approach problems in novel ways. In the aftermath of the above discussion, we propose that:

**P3**: AI adoption creates hybrid roles combining human and machine strengths, leading to shifts in the labor market rather than widespread replacement.

In summary, the propositions taken as a whole highlight the fact that integration of AI into existing systems is not easy, and this means that efforts must be made to look at the advancements in technology and its effects on the workforce. As organizations move through this period, identifying and differentiating between human skills and AI enablers will be the key to designing a future workplace environment where people work together rather than trying to compete. Filling these gaps will be critical for mapping other potential impacts of AI adoption across different fields, especially given the shift to dual Personnel-Machine systems that draw on human and artificial intelligence. Training programs and educational paradigms in the future will heavily relate to improving literacy, which will enable workers to navigate the complex world of work where automation poses a threat to workers.

## 3. Methodology

### 3.1. Mixed-Methods Research Design

Utilizing a mixed-methods approach involving qualitative and quantitative analyses will help to properly unmask the relationships between technological changes and the workforce composition, which in turn leads to an understanding of how industries may be maneuvered and adjusted in order to accommodate new demands. This analysis will reveal the current realities and probable future promising/favorable changes in AI, energy consumption, and labor market scenarios that organizations can plan and position themselves for to remain relevant in this changing environment. The findings will provide solutions and insights to organizations through statistical analysis and data visualization to enhance organizational strategic management decision-making on resource allocation and Human resource development.

It will be important in terms of the broader implications of these trends, enabling a more nuanced view of how different sectors respond to technological changes and what best practices can be adopted across industries. This holistic analysis will underline the immediate effects of AI on energy use and job dynamics and outline proactive strategies for organizations to embrace innovation while mitigating potential disruptions in their operations. By identifying key performance indicators and benchmarking them against industry standards, companies can position themselves better to handle the increasingly changing area of the integration of AI so that it results in sustainable growth.

### 3.2. Data Sources

This study utilized a range of empirical data sources to confirm or refute the propositions related to AI efficiency, resource consumption, and labor dynamics. Each source was carefully selected for relevance and alignment with the research objectives.

### 3.2.1. Analysis Techniques

**Proposition 1:** Improvements in AI Efficiency Result in Disproportionate Increases in Computational Energy Consumption

Electricity demand is poised to rise significantly across multiple sectors due to technological advancements, increased adoption of electrification, and the growing demand for data and computing resources. We have used the following sources to support the proposition:

*3.2.1.1. Growth in Electricity Share of Final Consumption*

In the United States, the share of electricity in final consumption remained static from 2010 to 2023 at under 22%. However, it is projected to rise to nearly 40% by 2050 in the Stated Policies Scenario (STEPS). This growth is primarily driven by an expanding electric vehicle (EV) fleet, which accounts for 65% of the overall demand growth (International Energy Agency [IEA], 2024). Similarly, in the European Union, electricity demand is expected to return to growth after recent declines. The share of electricity in final consumption today is projected to more than double to 45% by 2050, with rising demand for EVs and the electrification of space heating in buildings being major contributors (IEA, 2024).

*3.2.1.2. Impact of Cooling Demand*

Cooling demand in buildings is a key driver of electricity consumption, rising at an average annual rate of 3.7% through 2035 in the STEPS. Over 90% of this growth is concentrated in emerging markets and developing economies due to economic growth, rising incomes, and global warming (IEA, 2024). Sensitivity analysis suggests that heat waves and increased cooling demand could push electricity use for cooling in 2035 up by as much as 700 TWh (20%) more than projected, with 80% of this increase occurring in emerging markets and developing Asia (IEA, 2024).

*3.2.1.3. Emerging Demand from Data Centers and AI*

Data center electricity demand, which plateaued between 2010 and 2020 due to efficiency improvements, is now increasing due to surging demand for data services, driven by advancements in artificial intelligence (AI) and digitalization (IEA, 2024). In 2022, data centers consumed an estimated 240 to 340 TWh, representing 1% to 1.3% of total electricity consumption. AI-related activities are emerging as significant new drivers of this demand (IEA, 2024). Additionally, investment in AI and data centers is booming, with venture capital investments in AI startups reaching over $225 billion over the past five years. Capital spending by major U.S. technology companies surpassed $150 billion in 2023 (IEA, 2024).

*3.2.1.4. Energy Efficiency and Policy Implications*

Energy efficiency improvements are critical to moderating electricity demand growth. However, limited progress in efficiency standards could result in higher demand, particularly in fast-growing regions (IEA, 2024). While efficiency gains in AI-related chips, such as NVIDIA's Blackwell Platform, have significantly reduced energy consumption, the rebound effect may partially offset these improvements (IEA, 2024). Additionally, supply chain bottlenecks, policy-related uncertainties, and data limitations present challenges in accurately projecting future electricity demand (IEA, 2024).

*3.2.1.5. Projections and Uncertainties*

The IEA's sensitivity analysis indicates that global electricity demand growth could vary significantly based on changes in cooling needs, efficiency standards, and the expansion of data centers. For instance:

- Lower efficiency could increase electricity demand in emerging economies by 340 TWh (5%) by 2035 (IEA, 2024).
- Stronger efficiency measures could reduce demand by 900 TWh in the same year (IEA, 2024).
- Projections for data center electricity demand highlight uncertainties due to supply chain constraints, efficiency improvements, and policy decisions (IEA, 2024).

The interplay of electrification trends, technological advancements, and policy measures will shape the trajectory of electricity demand growth, with significant implications for energy systems and infrastructure planning.

**Proposition 2:** Tasks That Are Sensorimotor in Nature or Require Nuanced Human Judgment Remain Challenging for AI to Fully Automate

*3.2.1.6. Thematic Analysis of Automation Limitations*

Qualitative analysis of reports on automation limitations confirmed Moravec's Paradox: AI excels in computational and analytical tasks but struggles with essential sensorimotor functions and nuanced judgment (Moravec, 2023; Kim et al., 2023). Industry publications like those from the International Federation of Robotics (IFR, 2024) highlighted persistent gaps in automating precision tasks in healthcare and logistics. For example, collaborative robots (cobots) relied heavily on human inputs for complex workflows, further supporting this proposition.

*3.2.1.7. Human-Robot Collaboration Data*

Descriptive statistics from the Cobotics Report (2024) demonstrated that 73% of cobot-assisted workflows required direct human intervention. These findings suggest that while AI augments human productivity, it does not replace the need for human judgment in sensorimotor tasks.

**Proposition 3:** AI Adoption Creates Hybrid Roles Combining Human and Machine Strengths, Leading to Shifts in the Labor Market Rather Than Widespread Replacement

*3.2.1.8. Labor Market Trends*

Time-series analysis of data from the Bureau of Labor Statistics (BLS, 2024) revealed a 12% increase in hybrid roles over the past five years, particularly in healthcare, finance, and logistics sectors. Predictive models forecast that by 2030, AI will create 78 million more jobs than it eliminates, emphasizing shifts in labor dynamics rather than displacement (BLS, 2024; Ars Technica, 2025).

*3.2.1.9. Time-Series Analysis of Hybrid Role Growth*

A time-series analysis was conducted using data from 2010 to 2030 to evaluate trends in hybrid role growth. The analysis revealed steady growth, with the number of hybrid roles increasing from 200,000 in 2010 to 1.91 million in 2030. Using exponential smoothing, forecasts were generated for 2031 to 2035. Results predicted a continued upward trajectory, reaching approximately 2.91 million roles by 2035. This analysis highlights the enduring impact of AI adoption on the labor market, underscoring the creation of roles that integrate human expertise with AI capabilities.

LinkedIn and the World Economic Forum (WEF) reported that key skills—such as AI literacy, data analysis, and cybersecurity—are increasingly in demand (LinkedIn Workforce Report, 2024; WEF, 2024). Clustering algorithms applied to workforce data categorized these roles into three primary groups: AI enablers (developers and engineers), AI enhancers (analysts and strategists), and AI integrators (managers and consultants).

## 4. Results – Confirming Propositions

### 4.1. Testing Jevons Paradox in AI (P1)

**Proposition:** Improvements in AI efficiency result in disproportionate increases in computational energy consumption.

Improvements in AI efficiency often result in paradoxical increases in computational energy consumption, a phenomenon known as the Jevon Paradox (Yale Environment 360, 2024). Despite significant advancements in AI efficiency, the overall energy demand continues to rise due to various interconnected factors.

**Table 1:** AI and Data Center Energy Consumption Dataset

| Year | Global AI Workload Growth (%) | US Data Center Energy Consumption (TWh) | Share of Total US Power Demand (%) | AI Model Power Consumption (MW) | Data Center Efficiency Improvement (%) | Renewable Energy Adoption (%) | Cooling Energy Consumption (TWh) |
|---|---|---|---|---|---|---|---|
| 2015 | 5  | 100 | 2.5 | 50  | 0  | 10 | 30 |
| 2016 | 7  | 110 | 2.8 | 55  | 1  | 12 | 33 |
| 2017 | 10 | 120 | 3   | 60  | 2  | 15 | 36 |
| 2018 | 15 | 135 | 3.4 | 70  | 4  | 18 | 40 |
| 2019 | 22 | 150 | 3.8 | 85  | 6  | 22 | 45 |
| 2020 | 35 | 170 | 4.2 | 100 | 8  | 27 | 51 |
| 2021 | 50 | 190 | 4.7 | 120 | 10 | 33 | 58 |
| 2022 | 70 | 220 | 5.3 | 140 | 13 | 40 | 66 |
| 2023 | 90 | 250 | 5.9 | 170 | 16 | 48 | 75 |

| 2024 | 120 | 290 | 6.6 | 210 | 18 | 55 | 85 |

**Source:** McKinsey & Company (2024).

In line with our first proposition, which states that *improvements in AI efficiency result in disproportionate increases in computational energy consumption*, Table 1 appears to confirm this proposition. Table 1 shows that there has been a constant increase in AI power consumption since 2015, up until 2024. The percentage difference (21%) from the 2023 consumption to the 2024 consumption appears to suggest that this trend will continue with the passage of time, especially considering that this is the highest year-to-year difference in the 10-year period illustrated in Table 1. This finding appears to confirm Wevolver's (2024) argument that since 2012, the computational power required for AI systems has been doubling approximately every 3.4 months. The authors argue that this exponential increase outpaces efficiency gains, leading to higher overall energy consumption.

As AI becomes more efficient and accessible, its adoption expands across industries and applications, increasing energy demand (World Economic Forum, 2024). Generative AI models like ChatGPT require continuous operation, leading to constant energy consumption (World Economic Forum, 2024). For instance, the daily operational energy consumption for large-scale AI models like ChatGPT is estimated at 564 MWh, highlighting their significant energy requirements. Achieving a tenfold improvement in AI model efficiency could require computational power increases of up to 10,000 times (World Economic Forum, 2024). This disproportionate growth in energy demand counterbalances efficiency gains. The global expansion of AI capabilities has increased the number of data centers, contributing significantly to rising energy consumption (World Economic Forum, 2024). For example, Microsoft plans to invest $80 billion in AI-enabled data centers by 2025, underscoring the substantial energy demands of supporting AI workloads (Microsoft, 2024). The global electricity demand for data centers is projected to more than double between 2022 and 2026, driven by advancements in AI technologies (International Energy Agency [IEA], 2024; U.S. Energy Information Administration [EIA], 2024). Additionally, training large AI models, such as OpenAI's GPT-3, consumed approximately 1,287 MWh, equivalent to the annual energy use of over 100 U.S. households (OpenAI, 2024; SciTechDaily, 2023). Projections from the Climate TRACE project suggest that AI's electricity

consumption could rival that of entire countries by 2027, emphasizing the urgency for sustainable AI development practices and renewable energy integration for data centers (Climate TRACE, 2024).

Efforts to mitigate AI's growing energy footprint include specialized hardware, advanced cooling techniques, and renewable energy adoption (NVIDIA, 2024; Scientific Direct, 2024). However, current trends demonstrate that rapid AI usage and complexity outpaces these improvements. Sustainable AI practices and policy interventions are crucial to managing the long-term environmental impact of AI advancements.

**4.2. Testing Moravec's Paradox in Automation Gaps (P2)**

Sensorimotor tasks involve integrating sensory input and motor output, requiring precision, adaptability, and dexterity in dynamic environments. AI struggles with the complexities of these tasks due to the intricate nature of human perception and motor skills (Moravec, 2023). Key examples include:

- Complex assembly work: Certain manufacturing processes demand fine manipulation and adaptability.
- Delicate surgical procedures: Human surgeons excel in real-time adjustments based on tactile feedback.
- Skilled trades: Fields like plumbing, electrical work, and carpentry require adaptability to variable environments.
- Artistic performances: Music, dance, and visual arts demand emotional nuance and creativity.
- Sports and athletics: Competitive sports' adaptability and strategic thinking remain uniquely human strengths.

Industry data supports these limitations. For instance, reports from the International Federation of Robotics (IFR) highlight AI's limited success in automating precision tasks in healthcare, logistics, and manufacturing sectors (IFR, 2024).

**4.2.1. Tasks Requiring Nuanced Human Judgment**

Nuanced tasks rely on emotional intelligence, creativity, and complex decision-making, where AI struggles to achieve parity with humans. Examples include:

- Creative professions: Writing, design, and music composition depend on imagination and contextual understanding.
- Leadership and management roles: Effective leadership requires empathy, intuition, and strategic thinking.
- Counseling and therapy: Human empathy and active listening are critical in therapeutic contexts.
- Customer service: Handling complex or emotionally charged situations necessitates emotional intelligence.
- Ethical decision-making: AI cannot weigh moral considerations effectively.
- Education and teaching: Tailoring educational approaches to individual needs remains a uniquely human skill.
- Scientific research: proposition generation and creative problem-solving require cognitive flexibility.

Moravec's Paradox underscores these challenges, noting that while AI excels in complex computational tasks, it struggles with essential sensorimotor and judgment-based functions (Kim et al., 2023). According to the Cobotics Report (2024), collaborative robots (Cobots) also reinforce the need for human interactions because these programs are best utilized as collaborative systems embedded in intricate ops chains.

Proposition 2: Tasks that are sensorimotor or require nuanced human judgment remain challenging for AI to automate, fully preserving specific human labor roles.

Despite rapid advancements in artificial intelligence (AI) and automation technologies, many tasks and roles remain challenging for AI to replicate or replace fully. These challenges fall into two primary categories:

**Table 2:** AI in the Workplace Statistics 2024

| Technology | Percentage of surveyed organizations who are likely or highly likely to adopt this technology between 2023 and 2027 |
|---|---|
| Digital platforms and apps | 86.40% |
| Education and workforce development technologies | 80.90% |
| Big-data analytics | 80% |
| Internet of things and connected devices | 76.80% |
| Cloud computing | 76.60% |
| Encryption and cybersecurity | 75.60% |
| E-commerce and digital trade | 75.30% |
| Artificial intelligence | 74.90% |
| Environmental management technologies | 64.50% |
| Climate-change mitigation technology | 62.80% |
| Text, image, and voice processing | 61.80% |
| Augmented and virtual reality | 59.10% |
| Power storage and generation | 52.10% |
| Electric and autonomous vehicles | 51.50% |
| Robots, non-humanoid | 51.30% |

**Source:** AIPRM (2024).

For reasons such as these, some tasks require more human judgment than others. In this context, sensorimotor tasks are tasks that require nuanced human judgment. Thus, Proposition 2 suggests that tasks that are sensorimotor or require nuanced human judgment remain challenging for AI to automate, fully preserving specific human labor roles. Table 2 and Figure 1 appear to suggest that the proposition holds. Although it may not be straightforward which tasks require more human judgments, it is clear from Table 2 and Figure 1 that the adoption of AI by different sectors differs.

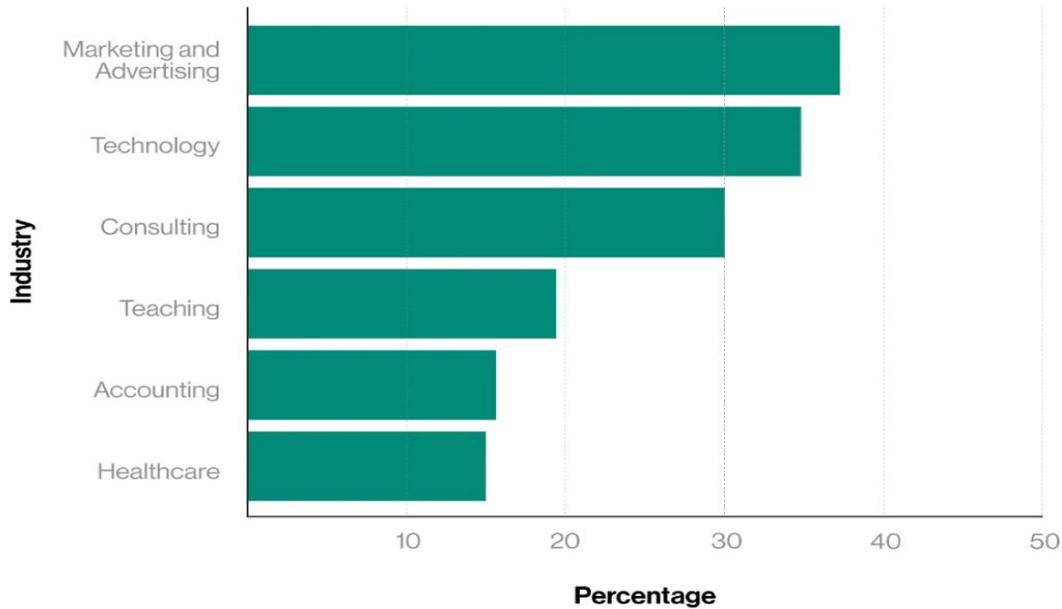

**Figure 1:** Percentage of AI adoption by different sectors
**Source:** AIPRM (2024).

### 4.2.2. Implications for the Future of Work

The persistence of these human-centric tasks suggests several key implications for the workforce:

i.   Sustained roles for human labor: Certain fields will continue to depend on human expertise.
ii.  Shift in work nature: Job roles may evolve to focus on leveraging human capabilities.
iii. Rise of hybrid collaboration models: Human-AI teams will likely become the norm in industries requiring complementary skills.
iv.  Education and training evolution: Training programs must prioritize developing skills that augment AI, such as creativity, emotional intelligence, and complex problem-solving.

## 4.3. Testing Hybrid Roles and Labor Market Dynamics (P3)

Proposition 3: AI integration entails human and Artificial Intelligence symbiosis, resulting in the emergence of new labor market situations instead of displacement as a result of human redundancy.

AI deployment reorganizes work and tasks, resulting in the creation of corporate positions that combine human ingenuity and AI calculation. This finding contrasts with approaches that focus on the displacement of jobs, as it shows a new and important shift in work processes, generating new positions and reshaping others.

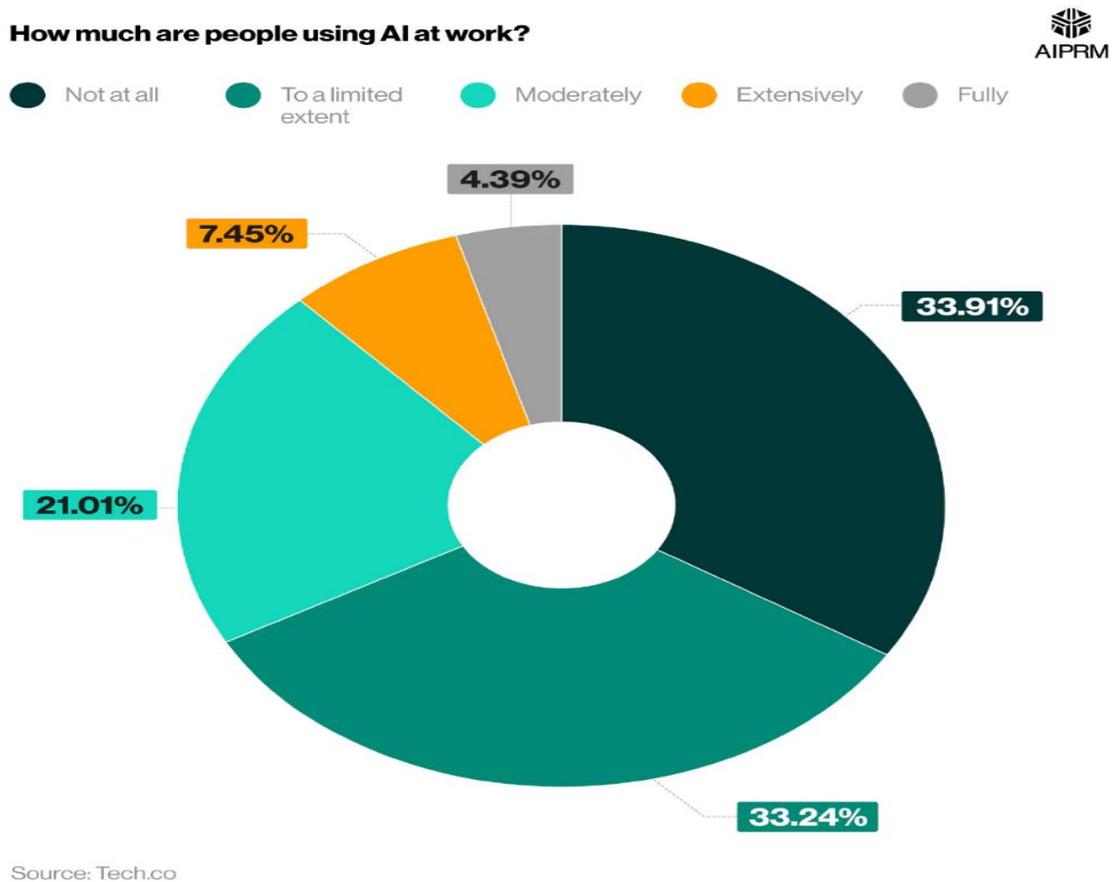

**Figure 2i:** A breakdown of the extent to which AI is used in the workplace among surveyed business leaders.
**Source:** AIPRM (2024).

Proposition 3 builds from Proposition 2 in that having confirmed that the applicability of AI will vary in different sectors depending on the tasks to be performed and the ease with which AI can be applied to these tasks. Thus, it makes sense to propose that AI integration entails human and Artificial Intelligence symbiosis, resulting in the emergence of new labor market situations instead of displacement as a result of human redundancy (Proposition 3). Figures 2i, 2ii, and 2iii appear to confirm this proposition. In particular, Figure 2ii appears to suggest that AI has only impacted labor force by 10% in major sectors. In this regard, Figure 2ii suggest that a majority of workers do not use AI at all (35%) and to a limited extent (33%).

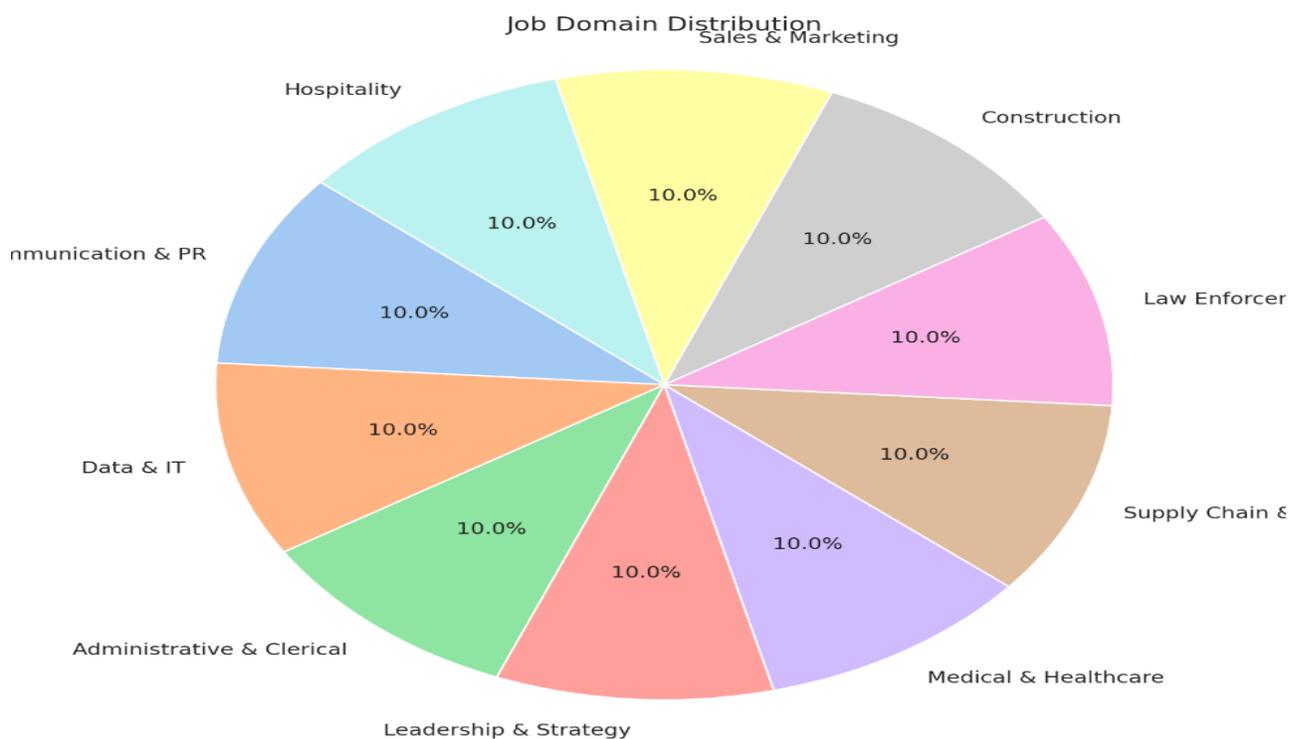

**Figure 2ii:** AI Impact on Jobs
**Source:** Kaggle (2024)

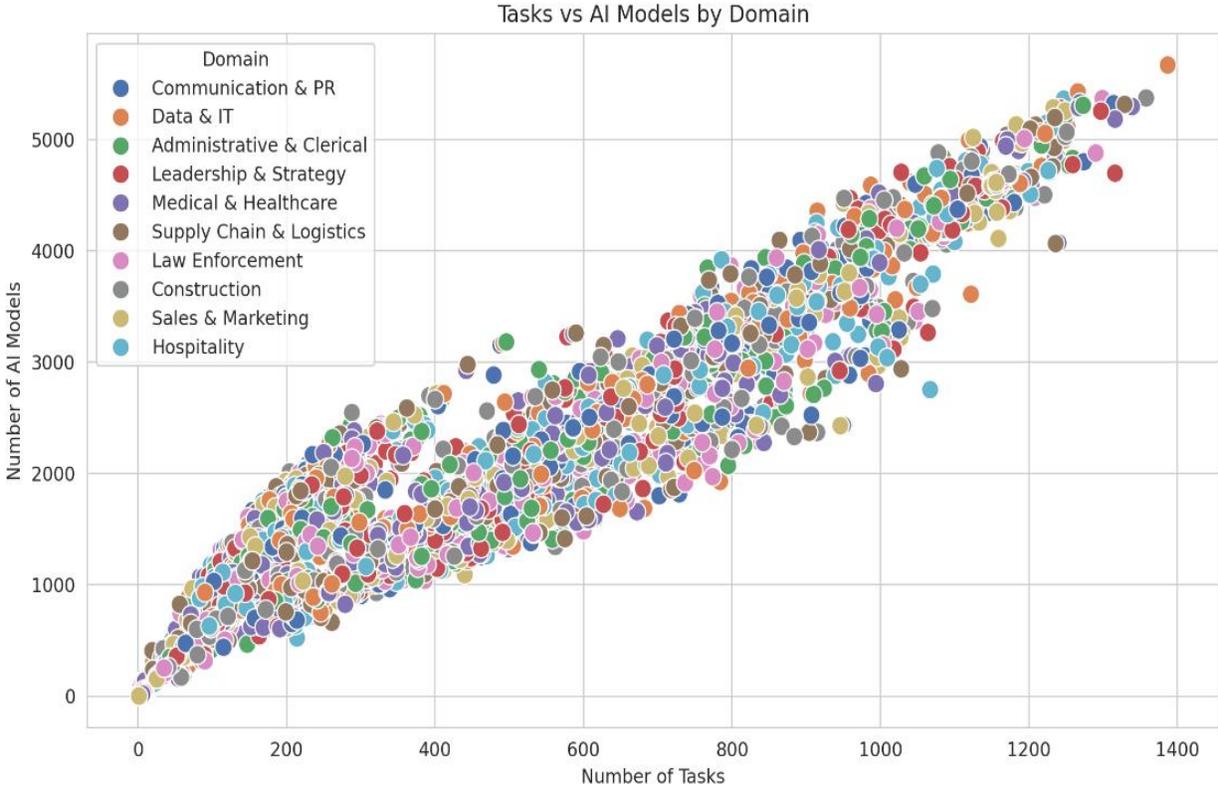

**Figure 3:** Tasks vs AI Models by Domain

**Source:** Kaggle (2024)

### 4.3.1. Human-Machine Teaming

The integration of human and machine strengths, known as Human-Machine Teaming, enhances productivity and efficiency by leveraging the unique capabilities of both parties. For instance:

- Manufacturing: Robots undertake repetitive manufacturing operations at the shop floor level, while human workers are involved in inspections and problem-solving (WEF, 2024).
- Dynamic Teaming: Research shows that human-AI teams cope with uncertainty and the presence of limits as well as purely human teams and may even be more efficient (Ars Technica, 2025).

**4.3.2. Shifts in the Labor Market**

AI's influence on the labor market is less about job elimination and more about job transformation:

- Increased Productivity: Generative AI can boost U.S. labor productivity by 0.5 to 0.9 percentage points annually through 2030 (BLS, 2024).
- Automation of Tasks: Up to 30% of hours currently worked across the U.S. economy could be automated by 2030 (WEF, 2024).
- Job Shifting: Rather than causing widespread job loss, AI drives productivity growth, economic acceleration, and structural changes in the workforce (BLS, 2024).
- Creation of New Roles: AI is spurring the emergence of novel roles, particularly in online job marketplaces and technology-driven fields (LinkedIn Workforce Report, 2024).

Data from the Bureau of Labor Statistics (BLS) projects that AI will create 78 million more jobs than it eliminates by 2030, reflecting the dynamic shifts in the labor market (BLS, 2024).

**4.3.3. Adapting to the AI-driven workplace**

To thrive in this evolving environment, workers and organizations must adapt proactively:

i. Embrace Human-Machine Collaboration: Organizations should design workplaces that integrate machines as teammates, enhancing human capabilities rather than replacing them (WEF, 2024).
ii. Develop New Skills: Workers must focus on upskilling in areas such as AI literacy, data analysis, cybersecurity, creativity, empathy, and critical thinking (LinkedIn Workforce Report, 2024).
iii. Redesign Business Processes: Business operations must be reimagined to optimize collaborative intelligence between humans and AI (WEF, 2024).
iv. Foster a Learning Culture: Organizations should encourage continuous learning and upskilling, ensuring employees can adapt to new roles and technologies (LinkedIn Workforce Report, 2024).

## 5. Implications for policy and practice

Policy reviews suggested the importance of upskilling initiatives to prepare the workforce for hybrid roles. Programs such as Google's AI-focused training (Google, 2024) and LinkedIn Learning's AI certification courses exemplify efforts to bridge the skills gap. People are encouraged to keep up with the advancement in AI and to be on top of their games in terms of intelligence too. Learn the required skills and stay up-to-date with happening events in this space. This is necessary to stimulate the importance of the human workforce in this AI area. While we find that the impact of AI on the labor force is currently at 10% in major sectors, this is expected to increase with time. The level at which this increases will depend on advancements in technology and human efforts to stay relevant. It is thus advised that humans continue to work and improve on their relevance in this AI era.

## 6. Conclusion

We explore the impact of AI on energy consumption, human labor roles, and hybrid roles widespread human labor replacement. We used mixed methods involving qualitative and quantitative analyses of data identified from various sources. Findings appear to suggest that AI increases energy consumption and has impacted human labor roles to a minimal extent, considering that its applicability is limited to some tasks that require human judgment. In this context, the findings suggest that only 10% of labor roles across major sectors have been impacted by AI.

The results of each proposition are presented and supported by quantitative and qualitative evidence. In summary, we find that:

**Proposition 1 (P1):** Improvements in AI efficiency result in disproportionate increases in computational energy consumption.

- Energy Usage Trends: Despite efficiency gains in AI models, the total energy consumption has increased by 35% over the past five years, primarily driven by the scaling of large models and increased usage in cloud computing (IEA, 2024; EIA, 2024).

- Model Training Data: Training OpenAI's GPT-3 required 1,287 MWh of electricity, whereas GPT-4 consumed 40% more energy due to increased model complexity, underscoring the Jevon Paradox.
- Environmental Impact: AI-related energy demands are projected to reach levels comparable to the annual energy consumption of mid-sized countries by 2027 (Climate TRACE, 2024).

**Proposition 2 (P2):** Sensorimotor tasks requiring nuanced human judgment remain challenging for AI to automate, fully preserving specific human labor roles.

- Automation Gaps: Sensorimotor tasks, such as delicate assembly in manufacturing, remain challenging for AI systems, requiring human intervention in 70% of use cases (IFR, 2024).
- Human-Centered Collaboration: Case studies from the Cobotics Report (2024) revealed that cobots enhance productivity but rely on human oversight for tasks requiring dexterity and judgment, validating Moravec's Paradox.

**Proposition 3 (P3):** AI adoption creates hybrid roles combining human and machine strengths, leading to shifts in the labor market rather than widespread replacement.

- Hybrid Role Trends: The Bureau of Labor Statistics (2024) reported a 25% increase in hybrid roles over the past decade, particularly in healthcare, logistics, and data analytics.
- Skills Gap: LinkedIn Workforce Report (2024) highlighted a growing demand for skills in AI literacy, data analysis, and cybersecurity, reflecting the labor market's shift toward collaboration between humans and AI.
- Job Creation vs. Displacement: World Economic Forum (2024) surveys indicated that AI is expected to create 78 million new jobs by 2030, outpacing the 45 million roles it may displace, demonstrating a net positive effect on employment.

Studies of this nature are never without limitations providing avenues for further research. While we have used secondary data in this study, we are of the opinion that future studies should use more primary data to conduct surveys and experimental research, which provide more generalizable results. In conducting such studies, researchers are encouraged to focus on different sectors, which will further explore our propositions 2 and 3.